\documentclass[reprint,aps,prl,twocolumn,superscriptaddress,preprintnumbers,letterpaper,natbib,floatfix,nofootinbib]{revtex4-1}
\usepackage{amsmath,amssymb,mathtools,bm}
\usepackage{graphicx, color}
\usepackage[dvipsnames]{xcolor}
\usepackage{float}
\usepackage{multirow}
\usepackage[colorlinks=true
,urlcolor=purple
,anchorcolor=blue
,citecolor=red 
,filecolor=magenta
,linkcolor=blue
,menucolor=blue
,linktocpage=true
,pdfproducer=medialab
,pdfa=true
]{hyperref}
\usepackage[utf8]{inputenc}
\usepackage[english]{babel}
\usepackage{soul}
\usepackage{overpic}
\usepackage{epsfig} 
\usepackage{url}
\usepackage{float}
\usepackage[export]{adjustbox}
\usepackage{cleveref}
\usepackage[lofdepth,lotdepth,caption=false,justification=RaggedRight]{subfig}

\newcommand{\es}[2] {\begin{equation} \label{#1} \begin{split} #2 \end{split} \end{equation}}
\newcommand{\al}[1]{\begin{align} #1 \end{align} }
\newcommand{\bea}{\begin{eqnarray}}
\newcommand{\eea}{\end{eqnarray}}
\newcommand{\mysec}[1]{\paragraph*{#1.}\!\!\!\!\!\!\!---}
\newcommand{\D}{{\rm d}}

\begin{document}
\title{Finite Bubble Statistics Constrain Late Cosmological Phase Transitions}
\author{Gilly Elor}
\affiliation{Weinberg Institute, Department of Physics, University of Texas at Austin, Austin, TX 78712, USA}

\author{Ryusuke Jinno}
\affiliation{Research Center for the Early Universe (RESCEU), Graduate School of Science, The University of Tokyo, Tokyo 113-0033, Japan}

\author{Soubhik Kumar}
\affiliation{Center for Cosmology and Particle Physics, Department of Physics,
New York University, New York, NY 10003, USA}
\affiliation{Berkeley Center for Theoretical Physics, Department of Physics, University of California, Berkeley, CA 94720, USA}
\affiliation{Theoretical Physics Group, Lawrence Berkeley National Laboratory, Berkeley, CA 94720, USA}

\author{Robert McGehee}
\affiliation{William I. Fine Theoretical Physics Institute, School of Physics and Astronomy,\\ University of Minnesota, Minneapolis, MN 55455, USA}
\affiliation{Leinweber Center for Theoretical Physics, Department of Physics,\\ University of Michigan, Ann Arbor, MI 48109, USA}

\author{Yuhsin Tsai}
\affiliation{Department of Physics, University of Notre Dame, IN 46556, USA}

\begin{abstract}\noindent

We consider first order cosmological phase transitions (PT) happening at late times, below Standard Model (SM) temperatures $T_{\rm PT} \lesssim$~GeV. The inherently stochastic nature of bubble nucleation and the finite number of bubbles associated with a late-time PT lead to superhorizon fluctuations in the PT completion time. We compute how such fluctuations eventually source curvature fluctuations with universal properties, independent of the microphysics of the PT dynamics. Using Cosmic Microwave Background (CMB) and Large Scale Structure (LSS) measurements, we constrain the energy released in a dark-sector PT. For 
0.1~eV~$\lesssim T_{\rm PT} \lesssim$~keV this constraint is stronger than both the current bound from additional neutrino species $\Delta N_{\rm eff}$, and in some cases, even CMB-S4 projections. Future measurements of CMB spectral distortions and pulsar timing arrays will also provide competitive sensitivity for keV~$\lesssim T_{\rm PT} \lesssim$~GeV.
\end{abstract}

\maketitle
\preprint{FTPI-MINN-23-20}
\preprint{UTWI-39-2023}

PTs have been studied extensively for decades in models of baryogenesis~\cite{Cohen:1990py,Fromme:2006cm,Hall:2019ank,Elor:2022hpa}, (asymmetric) dark matter~\cite{Cohen:2008nb,Zurek:2013wia,Baker:2016xzo,Hall:2019rld,Asadi:2021yml,Asadi:2021pwo,Hall:2021zsk,Elor:2021swj,Asadi:2022njl}, extended Higgs sectors~\cite{Profumo:2007wc, Chacko:2005pe,Schwaller:2015tja, Ivanov:2017dad}, and spontaneously broken conformal symmetry~\cite{Creminelli:2001th, Randall:2006py, Nardini:2007me, Konstandin:2010cd, Konstandin:2011dr, Baratella:2018pxi, Agashe:2019lhy, Agashe:2020lfz, Ares:2020lbt, Levi:2022bzt, Mishra:2023kiu}, among others. PTs may also generate gravitational waves (GW)~\cite{Kosowsky:1992rz,Kosowsky:1991ua,Kosowsky:1992vn,Kamionkowski:1993fg, Caprini:2015zlo,Caprini:2019egz,Caldwell:2022qsj,LISACosmologyWorkingGroup:2022jok} that can be observed in the near future. 

\emph{Late-time} PTs have garnered attention due to their possible connections to puzzling observations from PTAs~\cite{NANOGrav:2023gor, Antoniadis:2023rey} and the `$H_0$ tension': a discrepancy between the direct measurements of the Hubble constant $H_0$~\cite{Riess:2021jrx} and its value inferred from the CMB~\cite{Planck:2018vyg} (for a review see \cite{Schoneberg:2021qvd}). These PTs occur below SM temperatures of a GeV (redshift $z\sim 10^{13}$) and before matter-radiation equality ($z\approx 3400$). For instance, a PT around $z\approx10^4$ is motivated by the proposed New Early Dark Energy (NEDE) solution to the $H_0$ problem~\cite{Wetterich:2004pv,Doran:2006kp,Poulin:2018cxd,Niedermann:2019olb,Niedermann:2020dwg}. Since the Hubble tension favors such ``early-time'' solutions~\cite{Kamionkowski:2022pkx}, other ideas also use such PTs~\cite{Niedermann:2021vgd,Freese:2021rjq}. A PT at $z\sim 10^{10}$ has also been proposed as a source of the observed stochastic GW background measured by PTAs~\cite{NANOGrav:2023gor,NANOGrav:2023hvm,EPTA:2023fyk,Reardon:2023gzh,Xu:2023wog}. Even later PTs may ameliorate the cosmological constant problem \cite{Bloch:2019bvc}. 

Due to constraints from big bang nucleosynthesis (BBN) and the CMB, late-time PTs that occur entirely in a dark sector with no significant reheating to SM particles are favored~\cite{Bai:2021ibt}. Thus, we focus on PTs that only release GWs and other forms of dark radiation (DR). The gravitational backreaction on the SM sector is the only way to identify and constrain such dark-sector PTs. For larger couplings between the dark sector and SM, constraints stronger than our model-{\it independent} ones may exist. A well-known constraint on post-BBN dark PTs is the bound on the number of additional neutrinos, $\Delta N_{\rm eff} < 0.29$ at $95\%$ CL~\cite{Planck:2018vyg,Cielo:2023bqp}, derived from Baryon Acoustic Oscillations (BAO) and CMB measurements. This places an upper bound on the fraction of DR energy density compared to the total radiation energy density $f_{\rm DR}\equiv\rho_{\rm DR}/\rho_\text{tot}\lesssim 0.04$.

A PT proceeds via nucleation of bubbles of true vacuum inside the metastable phase. To estimate the typical number of bubbles, consider a comoving volume corresponding to multipole $\ell\sim 10^3$. If $T_\text{PT}\sim \text{ TeV}$, there are an enormous number of bubbles inside that comoving volume: $N_b \sim \left[a({\rm TeV}) H({\rm TeV})/(10^3 a_0 H_0)\right]^3 \sim \left[\tau_0/(10^3\tau({\rm TeV}))\right]^3 \sim 10^{34}$. Here $\tau({\rm TeV})$ and $\tau_0$ are the conformal times at $T={\rm TeV}$ and today, respectively, with the corresponding scale factors denoted by $a({\rm TeV})$ and $a_0$. However, if $T_\text{PT}\sim \text{ keV}$, for example, the number of bubbles is much less, $N_b \sim 10^{6}$.

In this Letter, we demonstrate that the finite number of bubbles involved in a PT gives rise to super-horizon perturbations in PT completion time $\propto 1/\sqrt{N_b}$ which eventually contribute to curvature perturbations. We present the first calculation of these perturbations using gauge-invariant observables. The curvature power spectrum on large length scales follows a universal power-law scaling in the PT parameters, independent of its microscopic details.\footnote{While previous studies have addressed the effect of a PT on curvature perturbations at a parametric level~\cite{Freese:2023fcr, Freese:2022qrl,Liu:2022lvz}, including the \emph{big bubble problem} in inflationary models~\cite{Copeland:1994vg}, they have not precisely determined the magnitude of the power spectrum.} Even for a dark PT with no direct coupling to the SM, CMB and LSS measurements constrain the resulting curvature perturbations. This in turn constrains $f_{\rm DR}$ even more than the current and projected $\Delta N_{\rm eff}$ limits when $T_\text{PT} \lesssim 1~\text{keV}$. Additionally, the power-law scale dependence of this new contribution to curvature perturbations can create distinct signatures in the CMB and matter power spectrum, enabling us to identify the origin of the perturbation as a late-time PT. 

\mysec{Super-horizon fluctuations in percolation time from number of bubbles}A PT proceeds through bubble nucleation and expansion; a point in space transitions to the true vacuum when a bubble engulfs it. However, this process is inherently {\it stochastic}, and therefore, not all points in space transition to the true vacuum at the same time. To quantify this, the bubble nucleation rate per unit time and volume is~\cite{hogan1983nucleation,Enqvist:1991xw,Hindmarsh:2015qta}
\es{eq:nucl}{
\Gamma = \Gamma_0 e^{-S(t)} \approx \Gamma_0 e^{-S(t_f)}e^{\beta(t-t_f)},
}
with $S$ the bounce action for nucleating a critical bubble and $\beta \approx -\D S(t_f)/\D t$. Here $t_f$ is a reference time to measure the progression of the PT. If the nucleation rate increases with time, the fraction of the Universe in the metastable phase can be determined as~\cite{Enqvist:1991xw},
\es{eq:h}{
h(t) = \exp\left(- e^{\beta(t-t_f)}\right).
}
This implies that for $\beta\gg H_{\rm PT}$, the Hubble scale at $t_f$, the PT completes very rapidly around the time $t_f$, within a small fraction of Hubble time. However, as mentioned above, the PT does not complete exactly at the same time everywhere. Given the factor of $\beta$ in Eqs.~\eqref{eq:nucl} and~\eqref{eq:h}, we expect the variance of the PT completion time to be inversely proportional to $\beta$, i.e., a slower PT with a small value of $\beta$ will exhibit a larger variance and vice versa.

To characterize the variation in the PT completion time, we denote the time at which a point $\vec{x}$ transitions to true vacuum by $t_c(\vec{x})$, and compute the two-point function $H_{\rm PT}^2\langle \delta t_c(\vec{x}) \delta t_c(\vec{y})\rangle$. $\delta t_c(\vec{x}) = t_c(\vec{x})-\bar{t}_c$ with $\bar{t}_c$ the average time of conversion (see the Supplementary Material for a detailed definition). We include factors of $H_{\rm PT}$ so that $H_{\rm PT} \delta t_c$ is dimensionless. Practically, it is easier to write $H_{\rm PT}^2\langle \delta t_c(\vec{x}) \delta t_c(\vec{y})\rangle = (H_{\rm PT} / \beta)^2 \times \beta^2 \langle \delta t_c(\vec{x}) \delta t_c(\vec{y})\rangle$ and calculate the last factor, thus separating the effect of cosmic expansion from the PT dynamics. It is more convenient to work with the dimensionless Fourier transformed two-point function denoted by:
\begin{align}\label{eq:Pdt_1}
\mathcal{P}_{\delta t} (k)
=
\frac{k^3}{2 \pi^2} \left(\frac{H_{\rm PT}}{\beta}\right)^2 \int \D^3 r~
e^{i \vec{k} \cdot \vec{r}}
\beta ^2 \langle \delta t_c (\vec{x}) \delta t_c (\vec{y}) \rangle,
\end{align}
where $\vec{r} = \vec{x} - \vec{y}$. Physically, this characterizes the correlation of $t_c$ between any two points separated by a distance $\sim 1/k$. $\mathcal{P}_{\delta t} (k)$ has two distinct contributions: single-bubble and double-bubble, corresponding to $t_c (\vec{x})$ and $t_c (\vec{y})$ being set by the same bubble or two bubbles (see the Supplementary Material).

$\mathcal{P}_{\delta t}(k)$ changes qualitatively for modes smaller or larger than the typical bubble size~$v_w/\beta$, where $v_w$ is the bubble wall velocity. For modes smaller than or comparable to the bubble size, we analytically compute $\mathcal{P}_{\delta t}(k)$ in the Supplementary Material assuming constant wall velocity. However, due to intricate fluid dynamics and magnetohydrodynamics effects, a translation between $\mathcal{P}_{\delta t}(k)$ and sourced curvature perturbations is involved and model-dependent. Therefore, the dependence of the curvature power spectrum on $k$ for $k_p\equiv k/a_{\rm PT}\gtrsim \beta/v_w$ is less universal and varies as the properties of the PT change. (Here $k_p$ is a physical wavenumber and $a_{\rm PT}$ is the scale factor at $t_f$.) On the other hand, scales $k_p \ll \beta/v_w$ have many bubbles contributing to the correlation function within a spatial volume of linear size $1/k_p$. Thus, $\mathcal{P}_{\delta t}$ is more universal and less sensitive to the details of the PT thanks to the central limit theorem.

We can understand the behavior of $\mathcal{P}_{\delta t}$ for $k_p \ll \beta/v_w$ as follows. The average separation between bubbles is~\cite{Enqvist:1991xw,Hindmarsh:2015qta}, $d_b \approx (8\pi)^{1/3} v_w/\beta$. In a given volume $V$, there are $N \sim V/d_b^3$ independent regions where bubble nucleation can take place in an uncorrelated fashion. As a result, the standard deviation in PT completion time, when averaged over this entire volume, scales as $1/\sqrt{N}$. Thus, given $N\sim 1/(k_p d_b)^3$ for a scale $k_p$, we expect ${\cal P}_{\delta t} \propto (k_p d_b)^3$. Also, the combination $\beta\times t$ is what appears in the nucleation rates in Eqs.~\eqref{eq:nucl} and~\eqref{eq:h}, so ${\cal P}_{\delta t} \propto 1/\beta^2$. Thus, for $k_p \ll \beta/v_w$ and $H_{\rm PT} \ll \beta$, the dimensionless power spectrum scales as
\es{eq:Pdt}{
{\cal P}_{\delta t} &\sim (8\pi) \left(\frac{H_{\rm PT}}{\beta}\right)^2 \left(\frac{v_w k_p}{\beta}\right)^3 \\
&= c (8\pi) v_w^3 (k\tau_{\rm PT})^3 \left(\frac{H_{\rm PT}}{\beta}\right)^5.
}
We have noted that $\tau_{\rm PT} = 1/(a_{\rm PT} H_{\rm PT})$ for PTs during radiation domination and introduced a constant prefactor $c$. From the qualitative arguments above we expect $c\sim 1$; a detailed calculation in the Supplementary Material gives $c\approx 2.8$. Another way to understand the $k^3$ scaling of ${\cal P}_{\delta t}$ for small $k$ is through Eq.~\eqref{eq:Pdt_1}. We show in the Supplementary Material that the correlation $\langle\delta t_c(\vec{x}) \delta t_c(\vec{y})\rangle \sim e^{-\beta r a_{\rm PT}/2}$ for $\beta r a_{\rm PT} \gg 1$. As a result for $k_p \ll \beta/v_w$, the exponential phase in  Eq.~\eqref{eq:Pdt_1} does not contribute, and the $k$-dependence of ${\cal P}_{\delta t}$ comes solely from the $k^3$ prefactor.

\begin{figure}[t]
\centering
    \includegraphics[width=8cm]{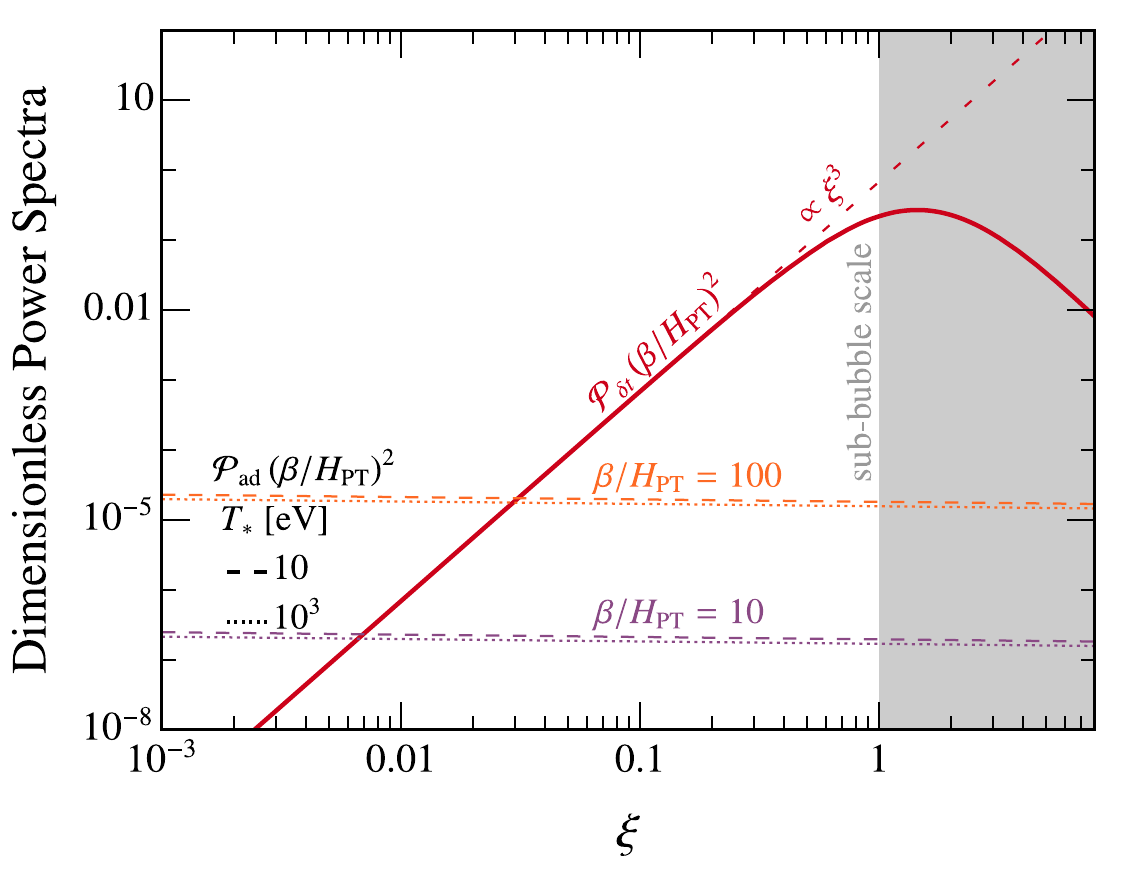}
    \caption{Dimensionless power spectrum of phase transition time fluctuation, re-scaled by $(\beta/H_{\rm PT})^2$ and plotted against comoving wavenumber ratio $\xi$ representing the perturbation mode relative to typical bubble separation. The PT spectrum (red) derived in the Supplementary Material is independent of $(\beta,\tau_{\rm PT})$, unlike adiabatic perturbations (purple, orange).}
    \label{fig:dmlsspowersp}
\end{figure}

The result for ${\cal P}_{\delta t}$ is in Fig.~\ref{fig:dmlsspowersp}, where $\xi \equiv k_p d_b = (8\pi)^{1/3} v_w (k\tau_{\rm PT})(H_{\rm PT}/\beta)$. For $\xi\ll 1$,  ${\cal P}_{\delta t} \sim \xi^3 (H_{\rm PT}/\beta)^2$, as expected from Eq.~\eqref{eq:Pdt}. However, close to $\xi\sim 1$, we see a deviation from that scaling. Below, we only use the result for ${\cal P}_{\delta t}$ for $\xi\leq 1$ since the regime $\xi\gg 1$ is sensitive to turbulence and magnetohydrodynamics effects. However, these sub-horizon inhomogeneities also give rise to density perturbations. In dark sectors in which sound waves dominantly source the GWs, the resulting constraints may even be stronger~\cite{Ramberg:2022irf}. 

We have described how the PT completion time fluctuates on superhorizon and `superbubble' scales. While such fluctuations are of `isocurvature' type initially (since they do not induce a change in energy density), eventually they source curvature perturbations.

\mysec{Curvature perturbations from fluctuations in percolation time}Consider two acausal patches, $A$ and $B$, where the PT takes place. Since our analysis relies on $\xi \sim 0.1-1$ and $\beta/H_{\rm PT} \lesssim 10^3$, we are in a regime where ${\cal P}_{\delta t} \gg A_s$, where $A_s =2.1 \times 10^{-9}$~\cite{Planck:2018jri} is the magnitude of the inflationary scalar power spectrum (we express ${\cal P}_{\delta t}$ in terms of gauge-invariant observables below). Thus, we can ignore effects due to $A_s$ and assume the PT takes place in a Universe that is a priori homogeneous in different patches. We will see how ${\cal P}_{\delta t}$ leads to inhomogeneities in the dark sector and how they then feed back into the SM sector, weighted by factors of $f_{\rm DR}$.  

Take the two patches $A$ and $B$ to each have size $v_w/\beta$ and equal energy density. $\rho_F$ is their (equal) false vacuum energy density and $t_c^A$ and $t_c^B$ their respective PT completion times. $t_c^A\neq t_c^B$ in general and we define the difference $t_c^B-t_c^A\equiv\delta t_c\ll t_c^{A,B} \sim 1/H_{\rm PT}$, with $\delta t_c>0$. When $A$ and $B$ undergo the PT, $\rho_F$ is converted into DR with an energy density $\rho_{\rm DR}$. Right at $t_c^A$, the energy densities in $A$ and $B$ are identical and the curvature perturbation is still zero. However, there is a non-zero isocurvature perturbation in DR at this time. This subsequently induces curvature perturbations as time evolves since DR and vacuum energy redshift differently. In other words, the equation of state of the Universe is not barotropic, i.e., the total pressure is not a definite function of the total energy, $p\neq p(\rho)$. As a result, the curvature perturbation is not constant (see e.g.~\cite{Garcia-Bellido:1995hsq, Wands:2000dp}) and evolves with time after the PT occurs.

We assume that (i) the PT takes place when the dark-sector energy density is dominated by the false vacuum and (ii) $\rho_F$ is entirely converted to $\rho_{\rm DR}$ after the PT. We can then write the DR energy density in the two patches at a later time $t_{\rm fin}$ as
\es{eq:rho_compare}{
\rho_{\rm DR}^{A,B}(t_{\rm fin}) = \rho_F \left(\frac{t_{c}^{A,B}}{t_{\rm fin}}\right)^2\,.
}
This shows that the energy densities of DR in the two patches are different and a nonzero DR density perturbation has been sourced by the DR isocurvature perturbation.\footnote{The different values of $\rho_{\rm DR}$ in $A$ and $B$ changes Hubble in the two patches, altering the energy-density redshift, but this correction is $\mathcal{O}(\delta\rho_{\rm DR}/\rho_{\rm SM})$ and negligible for our leading-order analysis.} We can compute this density perturbation using Eq.~\eqref{eq:rho_compare}, $\delta \rho_{\rm DR}/\rho_{\rm DR}=2\delta t_c/t_c$. Since we are working to leading order in perturbations, $\rho_\text{DR}\equiv \rho_{\rm DR}^{A,B}(t_{\rm fin})$ and likewise for $t_c$ in the denominators of the previous expression.

To compute the associated curvature perturbation, we can use the spatially flat gauge (for a review, see~\cite{Malik:2008im}), which amounts to comparing the energy densities in patches $A$ and $B$ at a common time $t_{\rm fin}$ when the scale factors are identical. Then the curvature perturbation (on uniform-density hypersurfaces) is
\es{eq:zeta_heu}{
\zeta =-\frac{H_{\rm PT}\delta\rho_{\rm DR}}{(\dot{\rho}_{\rm DR}+\dot{\rho}_{\rm SM})}&=\frac{1}{4}f_{\rm DR}\frac{\delta\rho_{\rm DR}}{\rho_{\rm DR}}
=\frac{1}{2}f_{\rm DR}\frac{\delta t_c}{t_c} \,.
}
So far, we have been assuming the dark-sector PT takes place when the dark sector is vacuum-energy dominated, i.e., the latent heat released in the PT is much larger than the temperature of the dark radiation bath. If not, we can include the standard parameter $\alpha_{\rm PT} = \rho_F/\rho_{\rm DR *} \lesssim 1$ where $\rho_{\rm DR *}$ is the energy density in the dark radiation bath at the percolation time. Including this factor, we arrive at the curvature power spectrum,
\begin{equation}\label{eq:Pzeta_tot_v1}
\mathcal{P}_\zeta(k)=\frac{1}{4}\left(\frac{\alpha_{\rm PT}}{1+\alpha_{\rm PT}}\right)^2 f_{\rm DR}^2\mathcal{P}_{\delta t}(k)+\mathcal{P}_{\rm ad}(k)\,.
\end{equation} 
In the last term, we have reintroduced the uncorrelated adiabatic perturbation $\mathcal{P}_{\rm ad}$. We take the pivot scale $k_*=0.05$~Mpc$^{-1}$, $A_s=2.1 \times 10^{-9}$, and tilt $n_s=0.966$~\cite{Planck:2018vyg} when calculating $\mathcal{P}_{\rm ad}$. The constraints on $\mathcal{P}_\zeta(k)$ can then be used to bound $f_{\rm DR}$ for different dark PT parameters. For an alternate derivation that relates ${\cal P}_{\delta t}$ to ${\cal P}_\zeta$ without relying on the separate Universe approach followed here, together with a derivation using the $\delta N$-formalism~\cite{Langlois:2008vk}, see the Supplementary Material.

Since we ignored the presence of inflationary, adiabatic perturbations while analyzing $\mathcal{P}_{\delta t}$, Eq.~\eqref{eq:Pzeta_tot_v1} is valid only for $\mathcal{P}_{\delta t}\gg {\cal P}_{\rm ad}$. In practice, given the current precision $\Delta{\cal P}_\zeta/{\cal P}_\zeta \sim 5\%$ on CMB scales, the above restriction puts an upper bound $f_{\rm DR} \lesssim 0.4$, above which the effects of $A_s$ would be relevant for determining ${\cal P}_{\delta t}$. Once the PT ends, all the dominant energy densities are in radiation, and superhorizon $\zeta$-modes remain constant until they enter the horizon. We note that PT also generates DR isocurvature, with a size roughly given by ${\cal P}_{\delta t}$, implying that isocurvature vanishes in the limit of ${\cal P}_{\delta t} \rightarrow 0$. The Planck constraint on DR isocurvature~\cite{Ghosh:2021axu} is similar to the constraints on curvature perturbation. Therefore, we will not consider the effect of isocurvature perturbations separately, but rather study their effects via the constraints on ${\cal P}_\zeta$.

\begin{figure}[t]
\centering
    \includegraphics[width=8.8cm]{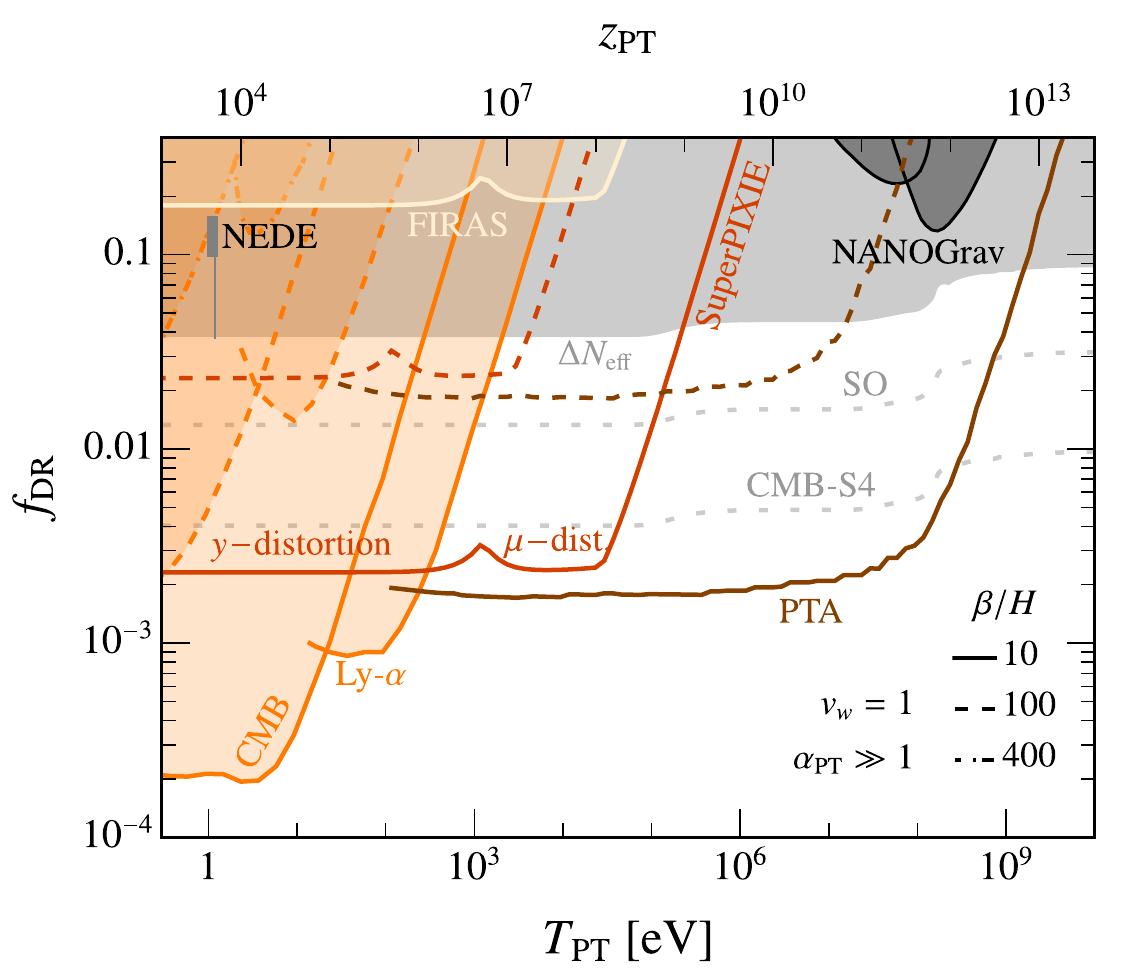}
    \caption{$2\sigma$ exclusion bounds on DR energy density fraction from current observations, as derived in this work. We show bounds using the CMB (Planck 2018~\cite{Planck:2018jri}) and Lyman-$\alpha$~\cite{Bird:2010mp} (pumpkin orange), as well as the FIRAS constraint~\cite{2002ApJ...581..817F} (candy corn yellow). The light grey region represents the existing bound $\Delta N_{\rm eff}\geq 0.29$~\cite{Planck:2018vyg}, and the dotted grey lines, the projected bounds from the Simons Observatory (SO) $\Delta N_{\rm eff} \ge 0.1$~\cite{SimonsObservatory:2018koc} and CMB-S4 $\Delta N_{\rm eff} \ge 0.03$~\cite{CMB-S4:2016ple}. Future projections from SuperPIXIE~\cite{Chluba:2012gq} (red), assuming sensitivity of $\Delta\rho_{\gamma}/\rho_\gamma\sim 10^{-8}$, and PTA~\cite{Lee:2020wfn} (maroon) are also depicted. We display the NEDE model's preferred region~\cite{Niedermann:2020dwg} (darker grey) and the PT generating the potential stochastic GW background~\cite{NANOGrav:2023hvm}. To assess existing NEDE model bounds, focus on $\beta/H_{\rm PT}$ values within the indicated bounds and disregard the grey region representing the $\Delta N_{\rm eff}$ bound.}
    \label{fig:fDRconst}
\end{figure}

\mysec{Time evolution of curvature perturbations}Perturbation modes with $k\tau_{\rm PT}\ll 1$ are outside the horizon when the PT takes place and we can characterize their subsequent cosmological evolution by just specifying ${\cal P}_\zeta(k)$. However, modes with $\xi\approx 1$ correspond to $k\tau_{\rm PT}\approx (8\pi)^{-1/3}(\beta/H_{\rm PT})v_w^{-1} \gtrsim 1$ for $\beta/H_{\rm PT}\gtrsim 10$, implying such modes are already inside the horizon when the PT takes place. To derive constraints based on ${\cal P}_\zeta(k)$ for a sub-horizon $k$-mode, we need to take into account that there is no sub-horizon evolution for a time $\Delta\tau\sim\tau_{\rm PT}-k^{-1}$, between mode reentry and the PT.

CMB temperature perturbations undergo diffusion damping while inside the horizon. A delay in subhorizon evolution by an amount $\Delta \tau$ implies PT-induced perturbations undergo less damping for a given $k$ compared to $\Lambda$CDM expectations. Starting with the same value of ${\cal P}_\zeta(k)$, the CMB anisotropies are larger in the PT scenario compared to $\Lambda$CDM. In this work, we take a conservative approach by not including this enhancement and leave a more precise computation for future work.

Perturbations in dark matter experience logarithmic growth in the radiation-dominated era upon horizon reentry. In $\Lambda$CDM cosmology, the power in a $k$-mode at the time of matter-radiation equality ($\tau_{\rm eq}\approx 110$~Mpc) is $\Delta_{\rm DM}(1/k,\tau_{\rm eq})^2{\cal P}_\zeta(k)$, where $\Delta_{\rm DM}$ denotes the matter transfer function: $\Delta_{\rm DM}(\tau_i,\tau)\approx 6.4\ln(0.44 \tau/\tau_i)$~\cite{Hu:1995en,dodelson:2003}. For the PT scenario, the analogous expression is $\Delta_{\rm DM}(\tau_{\rm PT},\tau_{\rm eq})^2{\cal P}_\zeta(k)$. We use a rescaled and weaker constraint ${\cal P}_\zeta(k) \times (\Delta_{\rm DM}(\tau_{\rm PT},\tau_{\rm eq})/\Delta_{\rm DM}(1/k,\tau_{\rm eq}))^2$ for $k>\tau^{-1}_{\rm PT}$ to take into account dark matter clustering bounds such as from Lyman-$\alpha$ and future PTA constraint on DM clustering.

\mysec{Cosmological Constraints}In Fig.~\ref{fig:fDRconst}, we present $2\sigma$ exclusion bounds on $f_\text{DR}$ using current constraints on $\mathcal{P}_\zeta$ and projected future sensitivities. We translate the comoving time $\tau_{\rm PT}$ in Eq.~\eqref{eq:Pdt} into the SM temperature and redshift at the time of the PT. CMB~\cite{Planck:2018jri} and Lyman-$\alpha$~\cite{Bird:2010mp} measurements set upper bounds on $\mathcal{P}_\zeta$ for $k$-modes up to $k\lesssim 3$~Mpc$^{-1}$. Our analysis excludes the pumpkin orange regions for various $\beta/H_{\rm PT}$ since in those regions, the PT contribution to ${\cal P}_\zeta$ is too large. Other constraints from ultracompact minihalos impacting PTAs may be relevant for $T_\text{PT} \gtrsim 1 \text{ MeV}$, but have unknown uncertainties related to the time of DM collapse~\cite{Clark:2015sha,Clark:2015tha}.

The $T_{\rm PT}$ dependence of the constraints in Fig.~\ref{fig:fDRconst} can be understood as follows. During a radiation-dominated epoch, $k_{\rm peak} \propto T_{\rm PT}$, where $k_{\rm peak}$ is the comoving wavenumber of the peak in ${\cal P}_{\delta t}(k)$. The constraint on $f_{\rm DR}$ for a given $T_{\rm PT}$ then depends on whether $k_{\rm peak}$ or the IR tail of ${\cal P}_{\delta t}(k)$ lies within the range probed by a given observable. Suppose that for a range of $T_{\rm PT}$, the corresponding range of $k_{\rm peak}$ is directly constrained by an observable. Then if the constraint on ${\cal P}_\zeta(k)$ over that range of $k_{\rm peak}$ is flat, the associated constraint on $f_{\rm DR}$ is also flat with respect to $T_{\rm PT}$, resulting in the plateaus in Fig.~\ref{fig:fDRconst}. This is because ${\cal P}_{\delta t}(k_{\rm peak})$ does not change as $T_{\rm PT}$ varies. This is what happens for the CMB bound for $T_{\rm PT}\lesssim 1$~eV for $\beta/H_{\rm PT}=10$. For larger $T_{\rm PT}$, $k_{\rm peak}$ lies outside the region probed by ${\cal P}_\zeta(k)$ constraints; constraints are only sensitive to the tail of the  ${\cal P}_{\delta t}$ distribution and $\propto f_{\rm DR}^2 k^3$ (from~\eqref{eq:Pdt} and~\eqref{eq:Pzeta_tot_v1}). In those regions, as $T_{\rm PT}$ is increased, the bound on $f_{\rm DR}$ goes as $1/T_{\rm PT}^{3/2}$ (since $T_{\rm PT} \propto k$). A similar transition from a plateau behavior is also seen at $\sim$100~eV for Ly-$\alpha$ and at $\sim$$100$~MeV for PTA. 

Notably, for $\beta/H_{\rm PT}\lesssim 400$, the bounds we derive from DR inhomogeneities are stronger than current $\Delta N_{\rm eff}$ constraints that track the homogeneous abundance of DR. For PTs that occur before BBN, there is a stricter bound of $\Delta N_{\rm eff} \lesssim 0.23$ when applying more observational constraints; for those PTs after, the constraint is slightly weaker at $\Delta N_{\rm eff} \lesssim 0.31$~\cite{Yeh:2022heq}. Since our new CMB$+$Ly-$\alpha$ bounds are in the latter range, and we want to show analogous $\Delta N_{\rm eff}$ constraints for CMB-S4, we plot Fig.~~\ref{fig:fDRconst} using the well-known $\Delta N_{\rm eff} < 0.29$~\cite{Planck:2018vyg}. For $\beta/H_{\rm PT}\sim 10$ and $T_{\rm PT}\lesssim$~keV, our analysis using CMB+Ly-$\alpha$ constrains such PTs as much as or better than the future Simons Observatory (SO)~\cite{SimonsObservatory:2018koc} and CMB-S4 projections on $\Delta N_{\rm eff}$~\cite{CMB-S4:2016ple}.

The NEDE model in~\cite{Niedermann:2019olb,Niedermann:2020dwg} favors $\alpha_{\rm PT}f_{\rm DR}$ in the upper (lower) dark gray region ($\pm1\sigma$) from the Planck18+BAO+LSS fit with (without) SH0ES data. While large values of $\beta/H_{\rm PT}$ generally require extra model-building, the model in~\cite{Niedermann:2020dwg} assumes $\beta/H_{\rm PT} \gtrsim 100$ and permits $\beta/H_{\rm PT}$ as large as $\sim10^3$ by including a field to trigger the PT. Still, our $\mathcal{P}_\zeta$ bound effectively disfavors the preferred NEDE region in $\left( T_\text{PT},f_\text{DR}\right)$ for all values of $\beta/H_{\rm PT} \lesssim$ 320 (230) with (without) SH0ES data.

For a large $\mathcal{P}_\zeta(k)$ with $k\lesssim 5400$~Mpc$^{-1}$ the PT can impact the dissipation of acoustic modes in photon-baryon perturbations, altering the photon's blackbody spectrum and inducing $\mu$- and $y$-distortions~\cite{Chluba:2012we,Hooper:2023nnl}: 
\begin{eqnarray}
X \!\! \simeq& \!\! A\displaystyle{\int_{k_{\rm min}}^\infty}\frac{\D k}{k}\mathcal{P}_\zeta(k)\left[B e^{-\frac{k}{5400/{\rm Mpc}}}-C e^{-(\frac{k}{31.6/{\rm Mpc}})^2}\right]
\end{eqnarray}
where $k_{\rm min}=1$~Mpc$^{-1}$,  $(A,B,C)_X=(2.2,1,1)_\mu$ and $(0.4,0,-1)_y$. Comparing this to the FIRAS bound of $|\mu|<9.0\times10^{-5}$ and $|y|<1.5\times 10^{-5}$ \cite{1994ApJ...420..439M,Fixsen:1996nj}, we derive the  exclusion bound labelled as `FIRAS'. When lowering $T_{\rm PT}$, the $y$-distortion bound takes over the $\mu$-bound around $T_{\rm PT}=10^3$~$(10^2)$~eV for the $\beta/H_{\rm PT}=10$~$(100)$ case. In contrast to Ref.~\cite{Liu:2022lvz}, our findings indicate that the FIRAS constraint is less stringent than the $\Delta N_{\rm eff}$ constraint, even for PT with small $\beta=10 H_{\rm PT}$.\footnote{Ref.~\cite{Liu:2022lvz} assumes the spatial energy density perturbation directly equals $\alpha_{\rm PT}f_{\rm DR}(\beta/H_{\rm PT})^{-5/2}$, omitting the numerical prefactor derived in the Supplementary Material. Their procedure in deriving $\mathcal{P}_\zeta(k)$ is also sensitive to the choice of a window function, while ours is not. Instead, our analysis derives $\mathcal{P}_\zeta(k)$ by starting with the primordial fluctuations and following the energy-momentum conservation equation for linear perturbations~\cite{Malik:2008im}. This could explain why our bounds differ.} 

Current $\mathcal{P}_\zeta$ measurements are less sensitive to PTs than the $\Delta N_{\rm eff}$ constraint for $T_{\rm PT}\gtrsim \text{ keV}$, but several proposed searches can constrain $\mathcal{P}_\zeta$ more powerfully and constrain weaker dark PTs. Super-PIXIE aims to measure the CMB with a sensitivity of $\Delta\rho_\gamma/\rho_\gamma\sim 10^{-8}$~\cite{Chluba:2019kpb} and the associated constraint is shown in red. PTAs can also probe $\mathcal{P}_\zeta$ by observing the phase shift in periodic pulsar signals mainly caused by the Doppler effect induced by an enhanced dark matter structure that accelerates Earth or a pulsar. The PTA sensitivity curves (maroon) use $\mathcal{P}_\zeta$ sensitivity derived in~\cite{Lee:2020wfn} that assumes 20 years of observations of 200 pulsars. This future sensitivity to PTs with $T_\text{PT} \gtrsim \text{ MeV}$ may test exotic DM models which rely on them~\cite{Elor:2021swj}. Also shown is the $2\sigma$-preferred PT region for the GW background hinted at by NANOGrav~\cite{NANOGrav:2023gor,NANOGrav:2023hvm} (darker grey, see, e.g.,~\cite{Franciolini:2023wjm} for alternative GW spectrum assumptions). At face value, this region conflicts with the $\Delta N_{\rm eff}$ constraint, but this prominent GW signal could largely originate from supermassive black hole mergers. With enhanced PTA measurements, we might still detect the PT signal within a comparable $T_{\rm PT}$ range. Then PTA measurements of $\mathcal{P}_\zeta$ could complement the GW detection.

\mysec{Discussion}We have demonstrated that finite bubble statistics can lead to superhorizon fluctuations in the PT completion time, regardless of the PT details. These fluctuations source curvature perturbations that affect the CMB, LSS, and other observables. Utilizing these, we find our constraints are in tension with some of the best fit regions of the NEDE models proposed to ameliorate the Hubble tension. At superhorizon scales, the (dimensionless) power spectrum of these fluctuations has a $k^3$-model-independent scaling since it is just determined by Poisson statistics. This contribution makes the {\it total} curvature perturbation scale non-invariant. Thus, the associated CMB phenomenology shares some similarities with the scale non-invariant effects due to `primordial features'~\cite{Chluba:2015bqa} produced during inflation and models with `blue-tilted' curvature perturbation~\cite{Kasuya:2009up, Kawasaki:2012wr, Chung:2021lfg, Ebadi:2023xhq}.

In our analysis, we have kept the $\Lambda$CDM parameters fixed. However, given the model-independent shape, one can do a joint analysis where both dark-sector and $\Lambda$CDM parameters are varied. We have also not considered constraints from modes that are smaller than typical bubbles as those are more model dependent. However, in the context of specific models one can obtain stronger constraints from such modes. We leave these for future work.

\emph{Acknowledgments.}~We thank Matthew Buckley, Zackaria Chacko, Peizhi Du, Anson Hook, Gordan Krnjaic, Toby Opferkuch, Davide Racco, Albert Stebbins, Gustavo Marques-Tavares, and Neal Weiner for helpful comments on the draft and discussions. The research of GE is supported by the National Science Foundation (NSF) Grant Number PHY-2210562, by a grant from University of Texas at Austin, and by a grant from the Simons Foundation. The work of RJ is supported by JSPS KAKENHI Grant Number 23K19048. SK is supported partially by the NSF grant PHY-2210498 and the Simons Foundation. RM was supported in part by DOE grant DE-SC0007859. YT is supported by the NSF grant PHY-2112540. RJ, SK, RM, and YT thank the Mainz Institute for Theoretical Physics (MITP) of the Cluster of Excellence PRISMA${}^+$ (project ID 39083149) for their hospitality while a portion of this work was completed. GE, SK, and YT thank Aspen Center for Physics (supported by NSF grant PHY-2210452) for their hospitality while this work was in progress.

\bibliography{refs}

\clearpage

\onecolumngrid
\begin{center}
  \textbf{\large Supplementary Material for Finite Bubble Statistics Constrain Late Cosmological Phase Transitions}\\[.2cm]
  \vspace{0.05in}
  {Gilly Elor, \ Ryusuke Jinno, \ Soubhik Kumar, \ Robert McGehee, \ and \ Yuhsin Tsai}
\end{center}

\setcounter{equation}{0}
\setcounter{figure}{0}
\setcounter{page}{1}
\makeatletter
\renewcommand{\theequation}{S\arabic{equation}}
\renewcommand{\thefigure}{S\arabic{figure}}

\section{Power spectrum of the transition time}

In this Supplementary Material we derive analytic expressions for the power spectrum of the transition time $t_c (\vec{x})$. The calculation is similar to Refs.~\cite{Jinno:2017fby,Jinno:2016vai}. We subtract the average transition time $\bar{t}_c$ and define $\delta t_c (\vec{x}) \equiv t_c (\vec{x}) - \bar{t}_c$, and calculate the power spectrum of the dimensionless quantity $\beta \delta t_c (\vec{x})$. It is handy to introduce the dimensionless power spectrum as usual
\begin{align}
{\cal P}_{\delta t}& \equiv {\cal P}_{\beta \delta t_c} (k)
=
\frac{k^3}{2 \pi^2} P_{\beta \delta t_c} (k),
\\
P_{\beta \delta t_c} (k)
&=
\int d^3 r~
e^{i \vec{k} \cdot \vec{r}}
\beta ^2 \langle \delta t_c (\vec{x}) \delta t_c (\vec{y}) \rangle
=
\int_0^\infty 4 \pi r^2 dr~
\frac{\sin (k r)}{k r} 
\beta ^2 \langle \delta t_c \delta t_c \rangle (r).
\label{eq:calP}
\end{align}
Though $\langle \delta t_c \delta t_c \rangle (r)$ depends only on $r\equiv|\vec{x}-\vec{y}|$, we sometimes use the notation $\langle \delta t_c (\vec{x}) \delta t_c (\vec{y}) \rangle$.

\begin{figure}
\begin{center}
    \includegraphics[width=0.35\columnwidth]{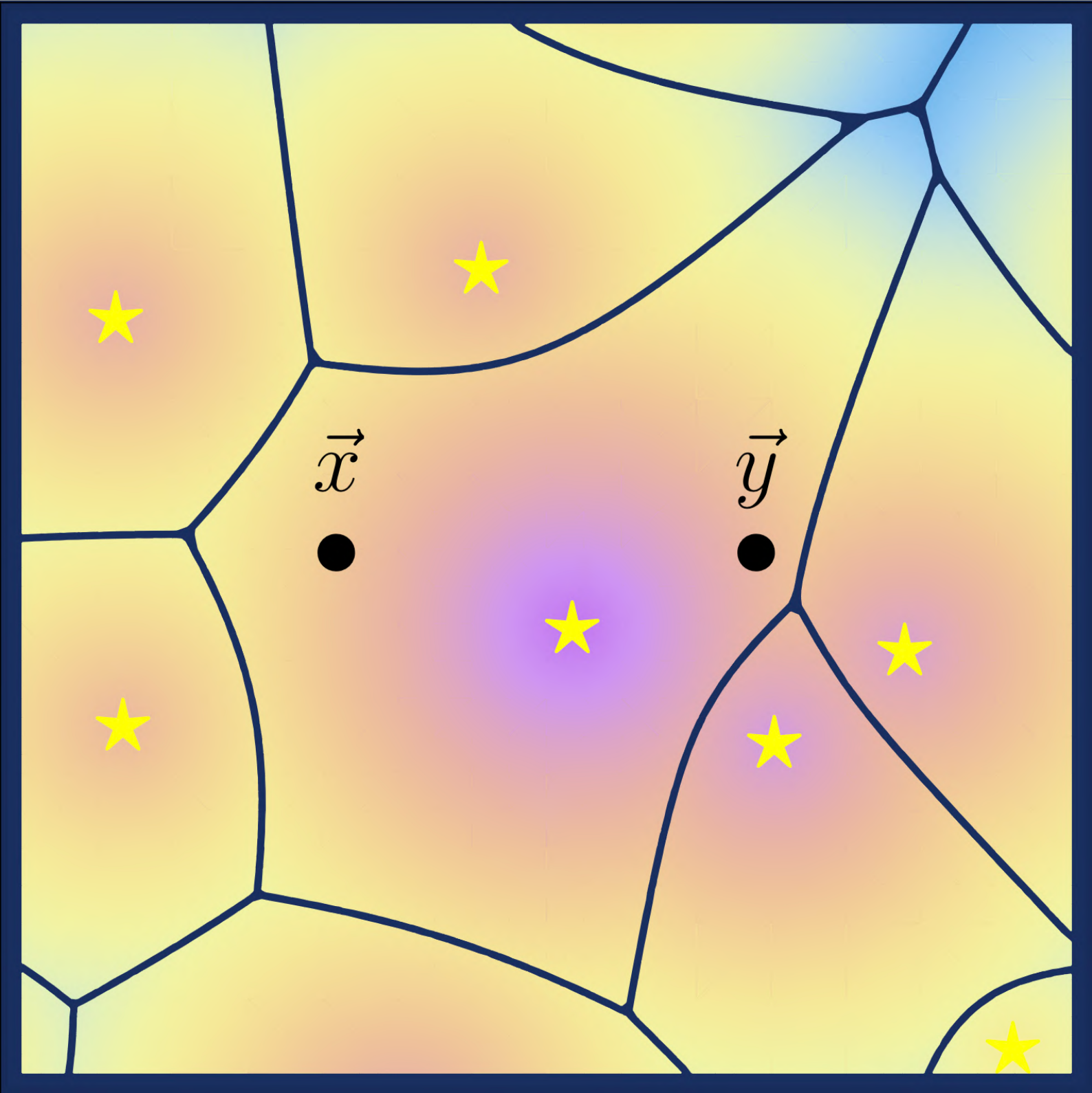}
    \hskip 1cm
    \includegraphics[width=0.35\columnwidth]{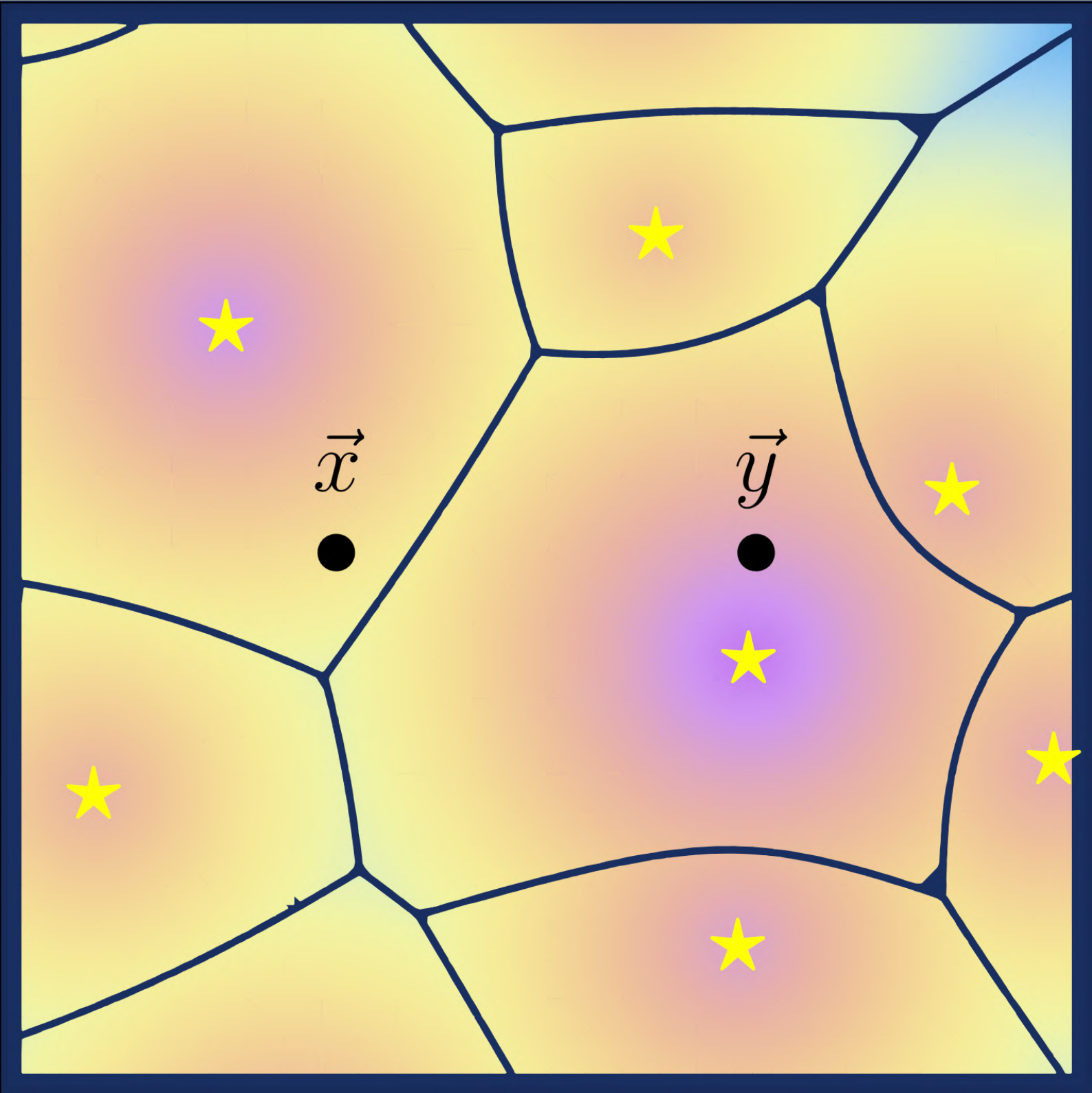}
    \caption{Two-dimensional illustration of the single-bubble (left) and double-bubble (right) contributions. The black dots are the evaluation points $\vec{x}$ and $\vec{y}$, while the stars denote the nucleation points of the bubbles. The color corresponds to the transition time for each spatial point. For the left case, the transition time at both points is determined by one single bubble nucleating around the center, while for the right case it is determined by two different bubbles nucleating at different locations.}
    \label{fig:single_double}
\end{center}
\end{figure}

Since we consider the two-point function of the transition time, we have two distinct contributions in $\langle \delta t_c (\vec{x}) \delta t_c (\vec{y}) \rangle$. They are illustrated in Fig.~\ref{fig:single_double}. We fix the evaluation points $\vec{x}$ and $\vec{y}$, and consider different realizations of bubble nucleation history. Since the transition time for each spatial point is determined by the bubble whose wall passes that point for the first time, the possibilities are that the transition times for the two points are determined by one single bubble, or that they are determined by two different bubbles. These two cases correspond to the left and right panels of Fig.~\ref{fig:single_double}.

\begin{figure}
\begin{center}
    \includegraphics[width=0.35\columnwidth]{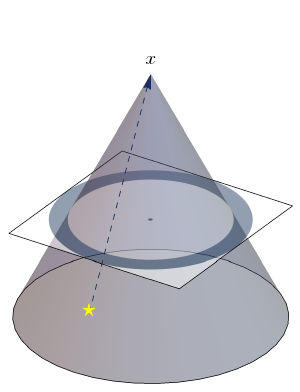}
    \includegraphics[width=0.55\columnwidth]{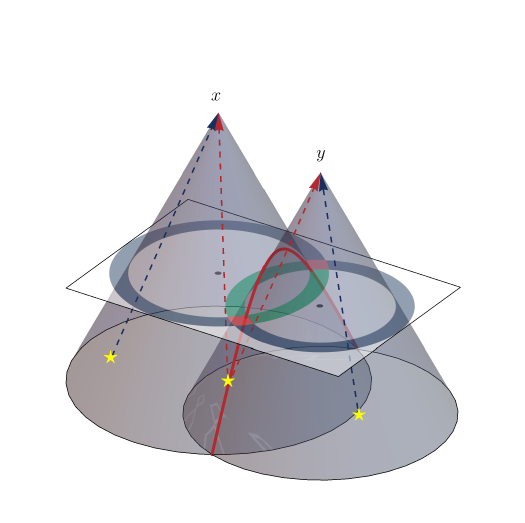}
    \caption{Past-cone geometry in 3D spacetime. Embedded are 2D spatial slices of past cones. Left: Past cone of $\vec{x}$. This geometry is used to calculate the average transition time $\bar{t}_c$. The survival probability $P (x)$ is defined as the probability for no bubble to nucleate inside the cone. A bubble nucleating on the surface of the cone passes $\vec{x}$ at the evaluation time $t_x$. Right: Past cones of $\vec{x}$ and $\vec{y}$. This geometry is used to calculate the two-point correlation function $\langle \delta t_c (\vec{x}) \delta t_c (\vec{y}) \rangle$. For the single-bubble case one single bubble nucleates at the intersection of the cone, while for the double-bubble two different bubbles nucleate on the surface of the each past cone.}
    \label{fig:bubblegeom}
\end{center}
\end{figure}

Let us consider the probability for $t_c (\vec{x})$ and $t_c (\vec{y})$ to be in the infinitesimal intervals $t_x < t_c (\vec{x}) < t_x + d t_x$ and $t_y < t_c (\vec{y}) < t_y + d t_y$, respectively. For this to happen, note that $\vec{x}$ and $\vec{y}$ must be in the false vacuum at $t_x$ and $t_y$. Thus we need (1) and ``(2-s) or (2-d)" in the following:
\begin{itemize}
\item[(1)]
$\vec{x}$ and $\vec{y}$ remain in the false vacuum at $t_x$ and $t_y$, respectively.
\item[(2)]
\begin{itemize}
\item[(2-s)]
One bubble nucleates at the intersection of the two past cones (red diamond in the right panel of Fig.~\ref{fig:bubblegeom}).
\item[(2-d)]
Two bubbles nucleate on the surface of the past cones, one being in the $\vec{x}$ side and the other in the $\vec{y}$ side (blue bands in the right panel of Fig.~\ref{fig:bubblegeom}).
\end{itemize}
\end{itemize}
Hereafter we assume luminal walls $v_w = 1$ for simplicity, and thus the cones are light-cones. Once we obtain the probability for these to happen, the correlator $\langle \delta t_c (\vec{x}) \delta t_c (\vec{y}) \rangle$ reduces to $\sum_{t_x} \sum_{t_y}$ (probability) $\times$ (value of $\delta t_x \delta t_y$). Since the two contributions are distinct, we may decompose the correlator into two terms
\begin{align}
\langle \delta t_c (\vec{x}) \delta t_c (\vec{y}) \rangle
&=
\langle \delta t_c (\vec{x}) \delta t_c (\vec{y}) \rangle^{(s)}
+
\langle \delta t_c (\vec{x}) \delta t_c (\vec{y}) \rangle^{(d)},
\label{eq:single_double}
\end{align}
with each being called the single- and double-bubble contribution. Note that the condition (1) is common to (2-s) and (2-d). In the following we use the nucleation rate
\begin{align}
\Gamma (t_n)
&=
\Gamma_* e^{\beta (t_n - t_*)},
\end{align}
with $\beta$ being constant and $t_*$ being the typical transition time.
Hereafter we adopt $\beta = 1$ unit.
We sometimes use a shorthand notation $x \equiv (t_x, \vec{x})$ and $y \equiv (t_y, \vec{y})$.

Now let us move on to the detailed calculation. We first calculate the average transition time $\bar{t}_c$. For a spatial point $\vec{x}$ to experience the transition between $[t_x, t_x, d t_x]$, it must be in the false vacuum at $t_x$. Such a probability, which we call ``survival probability'', is given by
\begin{align}
P_{\rm surv} (x)
&=
e^{- \int_{z \in ({\rm inside~the~past~cone~of~}x)} d^4 z~\Gamma (t_z)}
\nonumber \\
&=
\exp
\left[
- \int_{- \infty}^{t_x} d t_n~
4 \pi (t_x - t_n)^2
\Gamma_* e^{- (t_n - t_*)}
\right]
\nonumber \\
&=
\exp
\left[
- 8 \pi \Gamma_* e^{t_x - t_*}
\right].
\end{align}
Hence, the average transition time is obtained as
\begin{align}
\bar{t}_c
&=
\int_{- \infty}^\infty d t_x~
t_x
\int_{- \infty}^{t_x} d t_n~
4 \pi (t_x - t_n)^2
\Gamma (t_n)
P_{\rm surv} (x)
\nonumber \\
&=
t_* - \ln (8 \pi \Gamma_*) - \gamma,
\end{align}
with $\gamma$ being the Euler-Mascheroni constant.

We next calculate the probability for (1) to happen. Following a similar procedure as above, we get
\begin{align}
P_{\rm surv} (x, y)
&=
e^{- I (x, y)}
=
e^{- \int_{z \in ({\rm inside~the~past~cones~of~}\vec{x}{\rm ~and~}\vec{y})} d^4 z~\Gamma (t_z)}
=
e^{- \Gamma_* {\cal I} (x, y) e^{t_{\langle x, y \rangle} - t_*}},
\\
{\cal I} (x, y)
&=
8 \pi \left[
e^{t_{x, y} / 2}
+
e^{- t_{x, y} / 2}
+
\frac{t_{x, y}^2 - (r^2 + 4 r)}{4 r} e^{- r / 2}
\right].
\end{align}
Here we defined $t_{\langle x, y \rangle} \equiv (t_x + t_y) / 2$, $t_{x, y} \equiv t_x - t_y$, $t_{\langle x, y \rangle, n} \equiv t_{\langle x, y \rangle} - t_n$, $t_{\langle x, y \rangle, xn} \equiv t_{\langle x, y \rangle} - t_{xn}$, and $t_{\langle x, y \rangle, yn} \equiv t_{\langle x, y \rangle} - t_{yn}$. In the following we calculate the probability for ``(1) and (2-s)'' and that for ``(1) and (2-d)'', and estimate the correlator $\beta ^2 \langle \delta t_c (\vec{x}) \delta t_c (\vec{y}) \rangle$. The final expression is the sum of Eqs.~(\ref{eq:single}) and (\ref{eq:double}).

{\bf Single-bubble contribution.}
The single-bubble contribution corresponds to the cases in which one single bubble nucleates at the red diamond in the right panel of Fig.~\ref{fig:bubblegeom}. Noting that the red diamond forms a circle along the $\phi$ direction in three-dimensional space, we obtain
\begin{align}
&\langle \delta t_c (\vec{x}) \delta t_c (\vec{y}) \rangle^{(s)}
\nonumber \\
&=
\int_{|t_x - t_y| < r} d t_x d t_y~
(t_x - \bar{t}_c) (t_y - \bar{t}_c)
\int_{- \infty}^{(t_x + t_y - r) / 2} d t_n \int_0^{2 \pi} d \phi~
\frac{(t_x - t_n) (t_y - t_n)}{r} P_{\rm surv} (x, y) \Gamma (t_n)
\nonumber \\
&=
\int_{- \infty}^\infty d t_{\langle x, y \rangle}
\int_{- r}^r d t_{x, y}
\int_{r / 2}^\infty d t_{\langle x, y \rangle, n}~
\frac{2 \pi}{r}
e^{- \Gamma_* {\cal I} (x, y) e^{t_{\langle x, y\rangle} - t_*}} \Gamma_*
e^{t_{\langle x, y \rangle} - t_*} e^{- t_{\langle x, y\rangle, n}}
\nonumber \\
&\qquad \times
\left( t_{\langle x, y \rangle} + \frac{t_{x, y}}{2} - \bar{t}_c \right)
\left( t_{\langle x, y \rangle} - \frac{t_{x, y}}{2} - \bar{t}_c \right)
\left( t_{\langle x, y \rangle, n} + \frac{t_{x, y}}{2} \right)
\left( t_{\langle x, y \rangle, n} - \frac{t_{x, y}}{2} \right).
\end{align}
We first integrate out $t_{\langle x, y \rangle}$ and get
\begin{align}
\langle \delta t_c (\vec{x}) \delta t_c (\vec{y}) \rangle^{(s)}
&=
\int_{- r}^r d t_{x, y}
\int_{r / 2}^\infty d t_{\langle x, y \rangle, n}~
\frac{2 \pi e^{- t_{\langle x, y\rangle, n}}}{r \, {\cal I} (x, y)}
\left( t_{\langle x, y \rangle, n}^2 - \frac{t_{x, y}^2}{4} \right)
\left[
\left(
\ln \left(
\frac{{\cal I} (x, y)}{8 \pi}
\right)
\right)^2
-
\frac{t_{x, y}^2}{4}
+
\frac{\pi^2}{6}
\right],
\end{align}
and then integrate out $t_{\langle x, y \rangle, n}$ and get
\begin{align}
\langle \delta t_c (\vec{x}) \delta t_c (\vec{y}) \rangle^{(s)}
&=
\int_{- r}^r d t_{x, y}~
\frac{2 \pi e^{- r / 2}}{r \, {\cal I} (x, y)}
\left( \frac{r^2}{4} + r + 2 - \frac{t_{x, y}^2}{4} \right)
\left[
\left(
\ln \left(
\frac{{\cal I} (x, y)}{8 \pi}
\right)
\right)^2
-
\frac{t_{x, y}^2}{4}
+
\frac{\pi^2}{6}
\right].
\label{eq:single}
\end{align}

{\bf Double-bubble contribution.}
The single-bubble contribution corresponds to the cases in which two different bubbles nucleate on the surface of the past cones in Fig.~\ref{fig:bubblegeom}.
\begin{align}
&\langle \Delta t_c (\vec{x}) \Delta t_c (\vec{y}) \rangle^{(d)}
\nonumber \\
&=
\int_{|t_x - t_y| < r} d t_x d t_y~
(t_x - \bar{t}_c)
(t_y - \bar{t}_c)
\int_{- \infty}^{t_x} d t_{xn} \int_{- \infty}^{t_y} d t_{yn} \int d c_x \int d c_y \int_0^{2 \pi} d \phi_x \int_0^{2 \pi} d \phi_y~
\nonumber \\
&\qquad
(t_x - t_{xn})^2 (t_y - t_{yn})^2 P_{\rm surv} (x, y) \Gamma (t_{xn}) \Gamma (t_{yn})
\nonumber \\
&=
\int_{- \infty}^\infty d t_{\langle x, y \rangle} \int_{- r}^r d t_{x, y} \int_0^\infty d t_{x, xn} \int_0^\infty d t_{y, yn}~
16 \pi^2 e^{- \Gamma_* {\cal I} (x, y) e^{t_{\langle x, y\rangle} - t_*}} \Gamma_*^2 e^{2 t_{\langle x, y \rangle} - 2 t_*} e^{- (t_{x, xn} + t_{y, yn})}
\nonumber \\
&\qquad \times
t_{x, xn}^2 t_{y, yn}^2
\left( t_{\langle x, y \rangle} + \frac{t_{x, y}}{2} - \bar{t}_c \right)
\left( t_{\langle x, y \rangle} - \frac{t_{x, y}}{2} - \bar{t}_c \right)
f_x f_y.
\end{align}
Note that the integration ranges for $c_x$ and $c_y$ can be nontrivial since the blue bands in the right panel of Fig.~\ref{fig:bubblegeom} do not necessarily form complete circles (i.e., complete spheres in three dimensions). The fractions $f_x$ and $f_y$ in the last line take account of this
\begin{align}
f_x
&=
1
\quad \left( t_{x, xn} < \frac{t_{x, y} + r}{2} \right),
\quad
\frac{(t_{x, xn} + r)^2 - (t_{x, xn} - t_{x, y})^2}{4 r t_{x, xn}}
\quad \left( t_{x, xn} > \frac{t_{x, y} + r}{2} \right),
\\
f_y
&=
1
\quad \left( t_{y, yn} < \frac{- t_{x, y} + r}{2} \right),
\quad
\frac{(t_{y, yn} + r)^2 - (t_{y, yn} + t_{x, y})^2}{4 r t_{y, yn}}
\quad \left( t_{y, yn} > \frac{- t_{x, y} + r}{2} \right).
\end{align}
We first integrate out $t_{\langle x, y \rangle}$ and get
\begin{align}
\langle \delta t_c (\vec{x}) \delta t_c (\vec{y}) \rangle^{(d)}
&=
\int_{- r}^r d t_{x, y} \int_0^\infty d t_{x, xn} \int_0^\infty d t_{y, yn}~
\frac{16 \pi^2 e^{- (t_{x, xn} + t_{y, yn})}}{{\cal I}^2 (x, y)}
\nonumber \\
&\qquad \times
t_{x, xn}^2 t_{y, yn}^2
\left[
\left( \ln \left( \frac{{\cal I} (x, y)}{8 \pi} \right) - 1 \right)^2 - \frac{t_{x, y}^2}{4} + \frac{\pi^2}{6} - 1
\right]
f (x) f (y),
\end{align}
and then integrate out $t_{x, xn}$ and $t_{y, yn}$ and get
\begin{align}
\langle \delta t_c (\vec{x}) \delta t_c (\vec{y}) \rangle^{(d)}
&=
\int_{- r}^r d t_{x, y}~
\frac{16 \pi^2}{{\cal I}^2 (x, y)}
\nonumber \\
&\qquad \times
\left[
4
-
\frac{e^{- t_{x, y} / 2 - r / 2}}{2 r}
(r + t_{x, y} + 4) (r - t_{x, y})
-
\frac{e^{t_{x, y} / 2 - r / 2}}{2 r}
(r - t_{x, y} + 4) (r + t_{x, y})
\right.
\nonumber \\
&\qquad
\left.
+
\frac{e^{- r}}{16 r^2}
((r + 4)^2 - t_{x, y}^2) (r^2 - t_{x, y}^2)
\right]
\left[
\left( \ln \left( \frac{{\cal I} (x, y)}{8 \pi} \right) - 1 \right)^2 - \frac{t_{x, y}^2}{4} + \frac{\pi^2}{6} - 1
\right].
\label{eq:double}
\end{align}

\begin{figure}
\begin{center}
    \includegraphics[width=0.48\columnwidth]{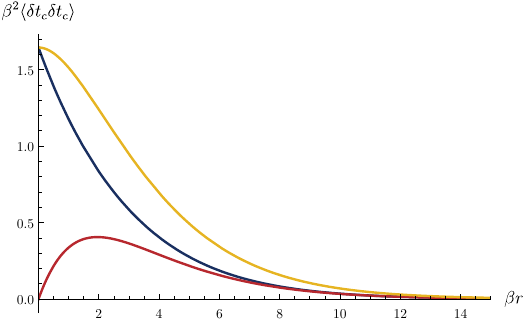}
    \caption{\small Single-bubble (blue) and double-bubble (red) contributions to the correlator $\beta^2 \langle \delta t_c (\vec{x}) \delta t_c (\vec{y}) \rangle$ and their sum (yellow).}
    \label{fig:deltcorrel}
\end{center}
\end{figure}

\begin{figure}
\begin{center}
    \includegraphics[width=0.48\columnwidth]{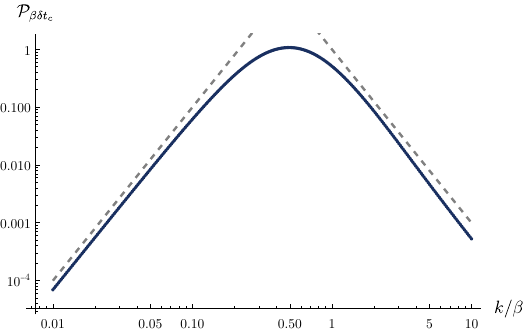}
    \caption{\small Power spectrum ${\cal P}_{\delta t} \equiv {\cal P}_{\beta \delta t_c}$. The dashed lines are ${\cal P}_{\beta \delta t_c} = 100 (k / \beta)^3$ and ${\cal P}_{\beta \delta t_c} = (k / \beta)^{-3}$ for comparison.}
    \label{fig:deltpowersp}
\end{center}
\end{figure}

{\bf Final expressions.}
The final expression for the correlator $\beta ^2 \langle \delta t_c (\vec{x}) \delta t_c (\vec{y}) \rangle$ is Eqs.~(\ref{eq:single}) and (\ref{eq:double}) substituted into Eq.~(\ref{eq:single_double}). These are one-dimensional integrals and easy to evaluate. Fig.~\ref{fig:deltcorrel} shows the two contributions and their sum. As expected, both decay exponentially at large $r$. Now it is straightforward to evaluate $P_{\beta \delta t_c}$. Fig.~\ref{fig:deltpowersp} shows $P_{\beta \delta t_c}$ calculated from Eq.~(\ref{eq:calP}). For small or large $k$, one may approximate the spectrum with ${\cal P}_{\beta \delta t_c} \simeq 70 (k / \beta)^3$ or ${\cal P}_{\beta \delta t_c} \simeq 0.7 (k / \beta)^{-3}$, respectively.

\section{Alternative Derivation of Curvature Perturbation}
\label{sec:der_eom}
Here we provide a derivation of~\eqref{eq:Pzeta_tot_v1} using the superhorizon evolution equation for the curvature perturbation $\zeta$.  On large scales, where spatial gradients can be ignored, $\zeta$ follows the equation~\cite{Garcia-Bellido:1995hsq, Wands:2000dp}
\es{eq:zeta_evol}{
\dot{\zeta} = - \frac{H}{\rho+p} \delta p_{\rm nad},
}
where
\al{
\delta p_{\rm nad} = \dot{p}\left(\frac{\delta p}{\dot{p}} - \frac{\delta \rho}{\dot{\rho}}\right),
}
and $H$ the Hubble rate, $\rho$ the total energy density and $p$ the total pressure. Now consider a dark sector PT that takes place at time $t_c$ that instantaneously converts the false vacuum energy into DR. This assumption of instantaneousness is justified since percolation takes place within a time window much narrower than a Hubble time, for $\beta \gg H_{\rm PT}$, as we consider. 

We can then write the total energy density as,
\al{
\rho = \rho_F \Theta(t_c - t) + \rho_F (a_c/a)^4 \Theta(t-t_c) + \rho_{\rm SM},
}
and
\al{
p = -\rho_F \Theta(t_c - t) + (\rho_F/3) (a_c/a)^4 \Theta(t-t_c) + \rho_{\rm SM}/3.
}
Here $\Theta(x)$ is the Heaviside theta function and $a_c$ is the scale factor at $t_c$. From these we can compute the time derivatives:
\al{
\dot{\rho} & = -\rho_F \delta(t_c - t) + (-4H)\rho_F (a_c/a)^4 \Theta(t-t_c) + \rho_F (a_c/a)^4 \delta(t-t_c) + \dot{\rho}_{\rm SM}, \\
& = (-4H)\rho_F (a_c/a)^4 \Theta(t-t_c) + \dot{\rho}_{\rm SM},\\
& \equiv B + \dot{\rho}_{\rm SM},
}
and
\al{
\dot{p} & = \rho_F \delta(t_c - t) + (-4H)(\rho_F/3) (a_c/a)^4 \Theta(t-t_c) + (\rho_F/3) (a_c/a)^4 \delta(t-t_c) + \dot{\rho}_{\rm SM}/3,\\
& = (4/3)\rho_F \delta(t_c - t) + (-4H)(\rho_F/3) (a_c/a)^4 \Theta(t-t_c) + \dot{\rho}_{\rm SM}/3,\\
& = (4/3)\rho_F \delta(t_c - t) + B/3 + \dot{\rho}_{\rm SM}/3.
}
We note $B \ll \dot{\rho}_{\rm SM}$ since $f_{\rm DR}\ll 1$ for our analysis. Using this we can approximate,
\al{
\frac{\dot{p}}{\dot{\rho}} = \frac{(4/3)\rho_F \delta(t_c - t) + B/3 + \dot{\rho}_{\rm SM}/3}{B + \dot{\rho}_{\rm SM}} \approx \frac{1}{3} +\frac{4\rho_F \delta(t-t_c)}{3 \dot{\rho}_{\rm SM}}.
}

So far we have computed the time derivatives of the homogeneous quantities $\rho$ and $p$ with time. To evaluate $\delta p_{\rm nad}$, we also compute the variations $\delta\rho$ and $\delta p$, which are variations in the total energy density and pressure, respectively, as we compare different acausal regions. As explained in the main text, we neglect inflationary fluctuations and therefore $\delta \rho_{\rm SM}=0$. However, since the PT does not complete everywhere at the same time, both $\delta t_c\neq 0$ and $\delta a_c \neq 0$. It is these variations that eventually source a non-zero curvature perturbation $\zeta$. The variation in the total energy density is given by,
\al{
\delta\rho & = \rho_F \delta(t_c - t) \delta t_c + (4 \delta a_c/a_c)\rho_F (a_c/a)^4 \Theta(t-t_c) + \rho_F (a_c/a)^4 \delta(t-t_c) (-\delta t_c) \\
& = (4 \delta a_c/a_c)\rho_F (a_c/a)^4 \Theta(t-t_c),
}
and in the total pressure by,
\al{
\delta p & = -\rho_F \delta(t_c - t) \delta t_c + 4(\delta a_c/a) (\rho_F/3) (a_c/a)^4 \Theta(t-t_c) + (\rho_F/3) (a_c/a)^4 \delta(t-t_c) (-\delta t_c), \\
& = -(4/3)\rho_F \delta(t_c - t) \delta t_c + 4(\delta a_c/a) (\rho_F/3) (a_c/a)^4 \Theta(t-t_c).
}
We now evaluate, to leading order in $f_{\rm DR}$,
\al{
\delta p - \frac{\dot{p}}{\dot{\rho}} \delta \rho \approx -\frac{4}{3}\rho_F \delta(t_c-t) \delta t_c.
}
Hence we can finally evaluate the change in $\zeta$ using~\eqref{eq:zeta_evol},
\al{
\dot{\zeta} & \approx -\frac{3 H}{4\rho_{\rm SM}} \left(-\frac{4}{3}\rho_F \delta(t_c-t) \delta t_c \right)\\
& \approx H \frac{\rho_F}{\rho_{\rm SM}} \delta(t_c-t) \delta t_c \\
& \approx \frac{\rho_F}{\rho_{\rm SM}} \delta(t_c-t) \frac{\delta t_c}{2t_c}.
}
Since the curvature perturbation before the PT is zero (since we assume inflationary fluctuations to be much smaller than the fluctuations induced by the dark PT), we can integrate the above to arrive at
\al{
\zeta = \frac{\rho_F}{\rho_{\rm SM}} \frac{\delta t_c }{2t_c}=f_{\rm DR} \frac{\delta t_c}{2t_c}~\text{(following the PT)}.
}
This matches with the expression in~\eqref{eq:zeta_heu}, obtained using the separate Universe approach.

For subhorizon modes, additional gradient terms would appear in Eq.~\eqref{eq:zeta_evol}~\cite{Malik:2008im}. However, those terms are homogeneous in $\zeta$ and the Newtonian potential, and therefore do not source $\zeta$ via $t_c$ fluctuations. Hence, we do not consider the gradient terms when evaluating how $\delta t_c$ sources $\zeta$. Once $\zeta$ is generated, however, we estimate the subhorizon evolution as mentioned in the main text.

\subsection{Derivation using the \texorpdfstring{$\delta N$}{d N} formalism}
\begin{figure}[t]
\centering
    \includegraphics[width=8.5cm]{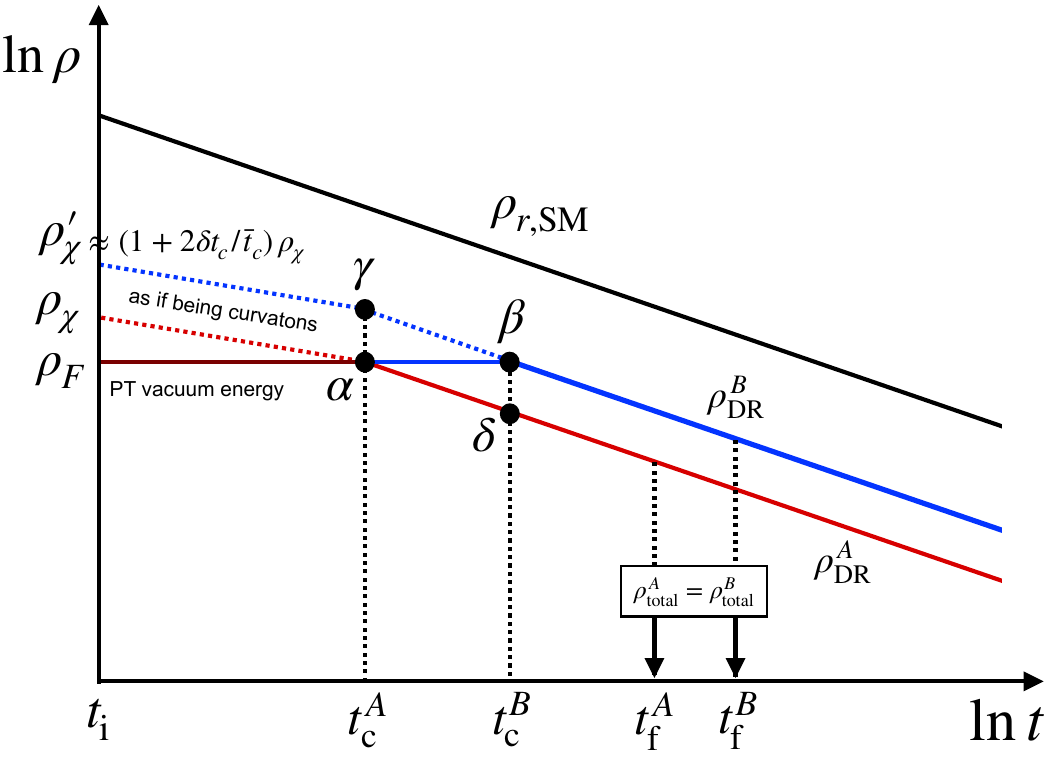}
    \caption{An illustration of the energy density evolution of dark PT. $\rho_{r,{\rm SM}}$ is the energy density of SM particles dominated by radiation. False vacuum energy of the two patches $A$ and $B$ reheats into DR at $t_c^{A,B}$. The total (DR and SM) energy density of the two patches are equal at $t_f^{A,B}$.}
    \label{fig:tPT}
\end{figure}
We can also derive the form of $\zeta$ by comparing the energy densities in the two patches $A$ and $B$, as discussed in the main text, and using the $\delta N$ formalism.
The evolution of the energy densities is shown schematically in Fig.~\ref{fig:tPT}.
The points $\alpha$ and $\beta$ denote the time of PT for patch $A$ and $B$, respectively.
At the point $\delta$, patch $A$ has a smaller energy density compared to patch $B$, as DR redshifts from point $\alpha$ to $\delta$.
This energy difference can be thought to arise from a hypothetical curvaton field.
In particular, we can translate the energy difference between $\beta$ and $\delta$, to an energy density difference at $t_c^A$, between $\alpha$ and $\gamma$.
Extrapolating to an even earlier time, we can imagine this difference to arise from {\it primordial} fluctuations of a curvaton field $\chi$, on a spatially flat hypersurface.
Given Eq.~\eqref{eq:rho_compare}, we find $\delta\rho_\chi = (2\delta t_c/t_c) \bar{\rho}_\chi$, where $\bar{\rho}_\chi$ is the homogeneous energy density in $\chi$.
The advantage of this approach is that we can now directly utilize the standard expressions derived in the curvaton scenario, see, e.g.,~\cite{Langlois:2008vk}.

We denote the energy densities of the SM radiation and $\chi$ by $\rho_{r,{\rm SM}}$ and $\rho_\chi$, respetively, and write them in terms of curvature perturbation on uniform density hypersurfaces for each species~\cite{Langlois:2008vk}:
\begin{equation}
\rho_{r,{\rm SM}}=\bar{\rho}_{r,{\rm SM}}e^{4(\zeta_{\rm inf}-\delta N)}\,,\quad \rho_{\chi}=\bar{\rho}_{\chi}e^{3(\zeta_{\chi}-\delta N)}\,.
\end{equation}
Here $\zeta_{\rm inf}$ is the curvature perturbation of the SM radiation (sourced by inflaton), and $\zeta_\chi$ is the curvature perturbation of $\chi$ right before its decay (sourced by PT time fluctuation).
On the uniform density hypersurface, the total radiation energy density,  $\rho_{r,{\rm SM}}+\rho_{\rm DR}\equiv\rho_r=\bar\rho_r\exp[4(\zeta_r-\delta N)]$, right after the curvaton decay equals $\bar\rho_r$, giving $\zeta_{r}=\delta N$. Given that $\rho_{\rm DR}=\rho_{\chi}$ right after the decay, we have
\begin{equation}
f_{\rm DR}e^{3(\zeta_\chi-\zeta_r)}+(1-f_{\rm DR})e^{4(\zeta_{\rm inf}-\zeta_r)}=1\,,\quad f_{\rm DR}=\frac{\bar{\rho}_{\rm DR}}{\bar{\rho}_{r,{\rm SM}}+\bar{\rho}_{\rm DR}}\,.
\end{equation}
The leading order expansion of the above equation in $\zeta_{\chi}$ and $\zeta_{\rm inf}$ gives
\begin{equation}\label{eq.zetar}
\zeta_r=\frac{3f_{\rm DR}}{4-f_{\rm DR}}\zeta_\chi+\frac{4(1-f_{\rm DR})}{4-f_{\rm DR}}\zeta_{\rm inf}\,.
\end{equation}
If $f_{\rm DR}\to 0$ ($f_{\rm DR}=1$), $\zeta_r=\zeta_{\rm inf}$ ($\zeta_r=\zeta_\chi$) as expected. In the post-inflation era, where the curvaton energy density  is very subdominant, the uniform energy density hypersurface is characterized by $\delta N=\zeta_{\rm inf}$, and the local density of $\chi$ is
\begin{equation}
\bar\rho_\chi e^{3(\zeta_\chi-\delta N)}=\bar\rho_\chi e^{3(\zeta_\chi-\zeta_{\rm inf})}=\bar\rho_\chi +\delta\rho_\chi\,.
\end{equation}
The leading order expansion gives
\begin{equation}
3\zeta_\chi=\frac{\delta\rho_\chi}{\bar\rho_\chi}+3\zeta_{\rm inf}=2\frac{\delta t_c}{\bar t_c}+3\zeta_{\rm inf}\,,
\end{equation}
and the total curvature perturbation
\begin{equation}
\zeta_r=\frac{2f_{\rm DR}}{4-f_{\rm DR}}\frac{\delta t_c}{\bar t_c}+\zeta_{\rm inf}\,.
\end{equation}
Since the PT perturbation is uncorrelated to the inflaton perturbation, we have for $f_{\rm DR}\ll 1$,
\begin{equation}
\mathcal{P}_{\zeta_r}\approx \frac{1}{4}f_{\rm DR}^2\mathcal{P}_{\delta t}+\mathcal{P}_{\rm ad}\,,
\end{equation}
where we have denoted the power spectrum of $\zeta_{\rm inf}$ by $\mathcal{P}_{\rm ad}$.
This reproduces Eq.~\eqref{eq:Pzeta_tot_v1} for $\alpha_{\rm PT} \gtrsim 1$.

\end{document}